\documentclass[aps,prd,twocolumn, superscriptaddress,nofootinbib]{revtex4}
\usepackage{fancyhdr}
\usepackage{wrapfig}
\usepackage{amsmath,amsthm,amssymb}
\usepackage{enumerate}
\usepackage{color}
\usepackage{graphicx}
\usepackage{caption}
\usepackage{subcaption}
\usepackage{sidecap}

\def\beq{\begin{equation}}
\def\eeq{\end{equation}}
\def\bea{\begin{eqnarray}}
\def\eea{\end{eqnarray}}

\def\bwt{\begin{widetext}}
\def\ewt{\end{widetext}}

\begin{document}

\title{Dynamical Emergence of Universal Horizons during the formation of Black Holes}

\author{Mehdi Saravani}\email{msaravani@pitp.ca}
\affiliation{Perimeter Institute
for Theoretical Physics, 31 Caroline St. N., Waterloo, ON, N2L 2Y5, Canada}
\affiliation{Department of Physics and Astronomy, University of Waterloo, Waterloo, ON, N2L 3G1, Canada}
\author{Niayesh Afshordi}\email{nafshordi@pitp.ca}
\affiliation{Perimeter Institute
for Theoretical Physics, 31 Caroline St. N., Waterloo, ON, N2L 2Y5, Canada}
\affiliation{Department of Physics and Astronomy, University of Waterloo, Waterloo, ON, N2L 3G1, Canada}
\author{Robert B. Mann}\email{rbmann@uwaterloo.ca}
\affiliation{Department of Physics and Astronomy, University of Waterloo, Waterloo, ON, N2L 3G1, Canada}
\affiliation{Perimeter Institute
for Theoretical Physics, 31 Caroline St. N., Waterloo, ON, N2L 2Y5, Canada}

\begin{abstract}
Motivations for the existence of a fundamental preferred frame  range from pure phenomenology to attempts to solve the non-renormalizability of quantum gravity, the problem of time (and scale), and the cosmological constant problem(s). In many explicit constructions, such as Einstein-Aether or Gravitational Aether theories, K-essence, Cuscuton theory, Shape Dynamics, or (non-projectable) Ho\v{r}ava-Lifshitz gravity, the low energy theory contains a fluid (which defines a preferred frame) with superluminal or incompressible excitations.  We study here the formation of black holes in the presence of such a fluid. In particular, we focus on the incompressible limit of the fluid (or Constant Mean Curvature foliation) in the space-time of a spherically collapsing shell within an asymptotically cosmological space-time. In this case, ignoring the fluid back reaction, we can analytically show that an observer inside $3/4$ of the Schwarzschild radius cannot send a signal outside, after a stage in collapse, even using signals that propagate infinitely fast in the preferred frame. This confirms the dynamical emergence of {\it universal horizons} that have been previously found in static solutions. We argue that this universal horizon should be considered as the future boundary of the classical space-time.  

\end{abstract}
\maketitle
\flushbottom

\section{Introduction}

General relativity (GR) has been the best classical theory of gravity that is compatible with a wide variety of experiments. 4D diffeomorphism invariance is the fundamental gauge symmetry of GR, resulting in the absence of a preferred frame. Despite this, there are good reasons to consider a fundamental preferred frame. Here, we list a few of the interesting alternative theories of gravity that invoke this property:
\begin{enumerate}

\item One reason to consider  theories with a preferred frame is purely phenomenological. An example is  Einstein-Aether theory \cite{Jacobson:2000xp}. The preferred frame is built into the theory via a unit time-like vector $u^\mu$. The action is the Einstein-Hilbert action plus all possible terms containing first order derivatives of $u^\mu$. This yields several free parameters that can be constrained/detected experimentally (e.g. \cite{Jacobson:2008aj,Yagi:2013qpa}). In particular, these constraints imply that aether disturbances should propagate (super)luminally \cite{Elliott:2005va}.

\item Another theory with a preferred frame is Gravitational Aether theory \cite{Afshordi:2008xu}, which is an attempt to solve the (old) cosmological constant problem by simply subtracting the trace of the energy-momentum tensor from the right hand side of Einstein's equations. This ensures that the zero point energy of quantum field theory does not gravitate. But in order to satisfy the Bianchi identities a new term (a symmetric tensor) must be added to the right hand side of Einstein's equations. The Bianchi identities then relate this term to the trace of matter energy-momentum tensor, via energy-momentum conservation. The new term is assumed to have the form of a perfect fluid (or the gravitational aether). In the limit of zero energy density (incompressibility), no new dynamical degree of freedom appears  and the Bianchi identities completely fix the evolution of the aether, whose four-velocity introduces a preferred direction of time.

\item Cosmological dynamical scalar fields generically introduce a preferred frame. An example of this type of theory is K-essence, constructed so that the scalar field develops a negative pressure once the matter dominated era begins. Its associated dynamical behaviour is then deemed responsible for the accelerating cosmic expansion of our universe, whilst avoiding both fine-tuning and anthropic arguments \cite{ArmendarizPicon:2000dh}.  However, any such model that solves such problems  necessarily has perturbations propagating faster than speed of light \cite{Bonvin:2006vc}.

\item  Pushing K-essence to its limit, cuscuton theory is a scalar field theory with {\it infinite} sound speed \cite{Afshordi:2006ad}. Cuscuton action is given by 
\beq
S=\int d^4x \sqrt{-g}(\mu^2\sqrt{\partial_{\nu}\phi\partial^{\nu}\phi}-V(\phi)).
\eeq
This theory is the same as the low energy limit of (non-projectable) Ho\v{r}ava-Lifshitz gravity for quadratic potential $V(\phi)$ \cite{Afshordi:2009tt}. The relation between parameters of cuscuton and $\lambda$ parameter of Ho\v{r}ava-Lifshitz gravity is as follows
\beq
\mu^2=-V''(\phi)=\frac{\lambda-1}{16\pi G_N}.
\eeq
They also have the same solution as Einstein-Aether theory when aether is hypersurface-orthogonal and $c_2=\lambda-1$ is the only non-vanishing term in Einstein-Aether action.

Constant field surfaces of cuscuton define a preferred time direction because signals propagate instantaneously on these surfaces. A constant field surface also has constant density and constant mean curvature. As a result, in a cosmological space-time, cuscuton can be considered as global time. 

\item Shape dynamics is an alternative theory of gravity whose fundamental symmetry is scale invariance \cite{Gomes:2010fh}. It has been shown  that shape dynamics and GR produce the same solutions in regions of space-time that admit a CMC slicing \cite{Gomes:2011zi}. Whether shape dynamics predicts a different solution (or even no solution) where there is no CMC slicing is still an open question.

\item Finally, Ho\v{r}ava-Lifshitz  gravity \cite{Horava:2009uw,Blas:2010hb} is a potentially renormalizable theory of gravity that breaks 4D diffemorphism invariant at high energies. However the non-projectable version of the theory reduces in the low energy limit to the Einstein-Hilbert action  together with a scalar field with infinite sound speed (cuscuton) \cite{Afshordi:2009tt}. 
\end{enumerate}

Spherically symmetric black hole solutions in Ho\v{r}ava-Lifshitz  gravity have been studied in \cite{Kehagias:2009is}, and it has been shown that the Schwarzschild metric is a solution to the equations of motion for large black holes (whose  curvature radius is much bigger than the Planck length). These solutions are the same as the spherically symmetric black hole solutions of Einstein-Aether theories \cite{Jacobson:2010mx}, since spherical symmetry requires the aether vector field to be hypersurface-orthogonal.
As long as the effect of the cuscuton on geometry is negligible, the Schwarzschild metric remains a black hole solution in Ho\v{r}ava-Lifshitz  gravity. However, the behavior of the cuscuton is important to the causal structure of spacetime, simply because its sound speed is greater than the speed of light.

In this paper, we investigate this issue.  Our motivation is to study  black hole solutions in theories with a preferred time direction. We specifically consider the causal structure of black hole solutions in Ho\v{r}ava-Lifshitz (or cuscuton) gravity.
Although spherically symmetric black holes in Ho\v{r}ava-Lifshitz (and Einstein-Aether) gravity are close to the Schwarzschild solution, they possess a new feature: they contain a trapped surface forbidding the escape of any signal, no matter how fast its propagation speed. This new type of horizon has been called a ``Universal'' horizon, as it is universal to all signals with arbitrary speed. Previously demonstrated for {\it static} spherically symmetric systems \cite{Blas:2011ni, Barausse:2011pu, Barausse:2012qh} (and \cite{Babichev:2006vx} for stationary solutions), we investigate here the collapse of a spherical thin shell and show how a universal horizon emerges in a dynamical setting. Also, unlike previous studies considering only asymptotically flat background, we have done our study of universal horizon in asymptotically cosmological solution.  We note that the dynamical formation of a similar additional trapped surface in K-essence models was also recently demonstrated, though there was numerical evidence of a breakdown of the initial value problem \cite{Akhoury:2011hr,Leonard:2011ce}.

 The structure of our paper is as follows. We start by   reviewing the propagation of a scalar field (with a general action) in a general background space-time. We then show how perturbations of the scalar field propagate through space-time, and  derive the ``propagation cone'' of perturbations at any given point. The propagation cone (sometimes called the sound cone) is an analogue of the light cone for the scalar field perturbations. For a scalar field, we find that the propagation cone depends on the constant background field surfaces. In Section III, we explicitly derive the equation of motion of background field and propagation cone in the limit where the sound speed is very large.  Section IV contains the solution for a collapsing spherical thin shell space-time.
We show that constant field surfaces are well behaved as long as the shell's radius ($R$) is bigger than Schwarzschild radius ($2M$). However, when $R$ approaches the critical value $R_c < 1.5M$ constant field surfaces start to stack up around the $r=1.5M$ surface. This behaviour shows that the field perturbations cannot escape from inside $r=1.5M$ to infinity even though they propagate almost instantaneously. As a result, there exists a horizon for these perturbations (universal horizon) at $r=1.5M$. This result is in agreement with the previous study of universal horizon in the infinite speed Einstein-Aether \cite{Berglund:2012bu}. Section V covers the emergence of universal horizon and  section VI concludes the paper.

\section{Introduction to Signal Propagation}\label{field}
Consider a scalar field $\phi$ with the following action
\beq \label{ee1}
S=\int d^4 x \sqrt{-g} \mathcal{L}\left(X,\phi\right),
\eeq
where $X=\frac{1}{2}g^{\mu \nu}\nabla_{\mu}\phi \nabla_{\nu}\phi$ and $g_{\mu \nu}$\footnote{metric signature $(+---)$} is the spacetime metric. We have restricted the Lagrangian $\mathcal{L}$ to depend only on the field and its first derivative. The energy-momentum tensor  
\beq\label{e2}
T_{\mu \nu}=\mathcal{L}_{,X}\nabla_{\mu}\phi \nabla_{\nu}\phi-g_{\mu \nu}\mathcal{L},
\eeq
is the same as that of a perfect fluid $T_{\mu \nu}=\left(\rho +p\right)u_{\mu}u_{\nu}-p~g_{\mu \nu}$ with
\bea
&u_{\mu}=\frac{\nabla_{\mu} \phi}{\sqrt{\nabla_{\alpha}\phi\nabla^{\alpha}\phi}},\label{e13}\\
&p=\mathcal{L},\label{e14}\\
&\rho=2X\mathcal{L}_{,X}-\mathcal{L}\label{e15},
\eea
provided that $X>0$ (so that the fluid has a rest frame). Henceforth we  assume that $X>0$, which has two advantages. Not only can the scalar field be understood as a perfect fluid (as noted above) but together with null energy condition (which requires that $\mathcal{L}_{,X} \ge 0$) this assumption implies that the spacetime is {\it stably causal} \cite{Babichev:2007dw}.

Variation of the action \eqref{ee1} with respect to $\phi$ yields the following equations of motion (for the derivation of the following equations \eqref{e7}-\eqref{e4} see \cite{Babichev:2007dw})
\beq\label{e7}
\tilde G^{\mu \nu}\nabla_{\mu}\nabla_{\nu}\phi+2X\mathcal{L}_{,X\phi}-\mathcal{L}_{,\phi}=0,
\eeq
where $\tilde G^{\mu \nu}=\mathcal{L}_{,X}g^{\mu \nu}+\mathcal{L}_{,XX}\nabla^{\mu}\phi \nabla^{\nu}\phi$.

To see how a $\phi$-signal propagates in this spacetime, consider a small field perturbation $\pi(x)$ on some background field $\phi_0(x)$ (neglecting the 
geometric back-reaction). These perturbations satisfy the following hyperbolic equation
\beq\label{e3}
\frac{1}{\sqrt{-G}}\partial_{\mu}\left(\sqrt{-G}G^{\mu \nu} \partial_{\nu}\pi\right)+M_{eff}^2\pi=0,
\eeq
where
\bea
&G^{\mu \nu}=\frac{c_s}{\mathcal{L}_{,X}^2}\tilde G^{\mu \nu},~\left(G^{-1}\right)_{\mu \nu}G^{\nu \rho}=\delta_{\mu}^{\rho},\\
&\sqrt{-G}=\sqrt{-det\left(G^{-1}\right)_{\mu \nu}},\\
&M_{eff}^2=\frac{c_s}{\mathcal{L}_{,X}^2}\left(2X\mathcal{L}_{,X\phi \phi}-\mathcal{L}_{,\phi \phi}+\frac{\partial \tilde G^{\mu \nu}}{\partial \phi}\nabla_{\mu}\nabla_{\nu}\phi_0\right),\\
&c_s^2=\frac{1}{1+2X\frac{\mathcal{L}_{,XX}}{\mathcal{L}_{,X}}}.\label{ee2}
\eea
The quantity $c_s$ is the propagation speed of the field perturbation $\pi$ in the field rest frame (co-moving frame).

Equation \eqref{e3} is a Klein-Gordon equation with the effective metric
\beq\label{e4}
\left(G^{-1}\right)_{\mu \nu}=\frac{\mathcal{L}_{,X}}{c_s}\left(g_{\mu \nu}-c_s^2\frac{\mathcal{L}_{,XX}}{\mathcal{L}_{,X}}\nabla_{\mu} \phi_0 \nabla_{\nu}\phi_0\right),
\eeq
which determines the propagation of perturbations. Indeed, $\left(G^{-1}\right)_{\mu \nu}$ defines a ``propagation" cone at any point in spacetime, through 
\beq
\left(G^{-1}\right)_{\mu \nu}v^{\mu}v^{\nu}=0.
\eeq

Using the above equations, we get
\beq
g_{\mu \nu}v^{\mu}v^{\nu}=\left(1-c_s^2\right)\left(g_{\mu \nu}u^{\mu}v^{\nu}\right)^2 
\eeq
showing that for  superluminal perturbations ($c_s>1$), the vector $v$ must be space-like with respect to the metric $g_{\mu \nu}$, consistent with our expectation that the influence cone is wider than the light cone for superluminal propagation. It also shows that the propagation cone at any point depends on the background field through the vector field $u^{\mu}$.

From now on we will focus on a scalar field with the following Lagrangian
\beq\label{e5}
\mathcal{L}=a X^n-V(\phi),
\eeq
where $a$ and $n$ are constants. So, Equation \eqref{ee2} yields
\beq
c_s^2=\frac{1}{2n-1}.
\eeq
Note that $\frac{1}{2}<n<1$ and $n>1$ respectively correspond to superluminal and subluminal propagation. The fluid also becomes incompressible (i.e. infinite speed of sound) at $n=\frac{1}{2}$. For $n<\frac{1}{2}$ the sound speed becomes imaginary, which is a sign of instability. We are interested in the superluminal case $\frac{1}{2}<n<1$. The propagation cone is then given by
\beq\label{e6}
g_{\mu \nu}v^{\mu}v^{\nu}=\frac{2\left(n-1\right)}{2n-1}\left(g_{\mu \nu}u^{\mu}v^{\nu}\right)^2.
\eeq
The right hand side of equation \eqref{e6} is negative for $\frac{1}{2}<n<1$, implying $v$ is space-like. Normalizing  $g_{\mu \nu}v^{\mu}v^{\nu}= -1$,  we get
\beq\label{e8}
\left(u_{\mu}v^{\mu}\right)^2=-\frac{2n-1}{2\left(n-1\right)}.
\eeq
Note that as $n$ approaches $\frac{1}{2}$, $v$ becomes orthogonal to the velocity vector $u$. It means that the propagation cone becomes almost tangent to constant field surfaces, for which perturbations propagate (almost)  parallel to the constant background field surfaces.

In summary, in order to determine how $\phi$-signals (perturbations) propagate through spacetime, we first solve Equation \eqref{e7} for the background field $\phi_0$. Equation \eqref{e8} then determines the propagation cone at any point of spacetime. Henceforth, we shall focus on fields with a very large sound speed, for which the values of $n$ are very close to $\frac{1}{2}$.
\section{Background Field and Propagation Cone}
Equation \eqref{e7} (together with \eqref{e5}) yields
\beq\label{e9}
anX^{n-1}\left(g^{\mu \nu}+(n-1)\frac{\nabla^{\mu}\phi\nabla^{\nu}\phi}{X}\right)\nabla_{\mu}\nabla_{\nu}\phi+V'(\phi)=0.
\eeq
where $n=\frac{1}{2}(1+\epsilon^2)$, $\epsilon \ll 1$. In order to find the propagation cone for small perturbations, we need to solve \eqref{e8} and \eqref{e9}, which we shall do as a power series in $\epsilon$. Considering first the equations for $n=\frac{1}{2}$ (zeroth order in $\epsilon$), we have
\bea
&\frac{a\sqrt{2}}{2}\nabla_{\mu}u^{\mu}+V'(\phi)=0,\label{e10}\\
&u_{\mu}v^{\mu}=0,
\eea
where we have used 
\beq
\nabla_{\mu}u^{\mu}=\frac{1}{\sqrt{\nabla_{\alpha}\phi\nabla^{\alpha}\phi}}\left(g^{\mu \nu}-\frac{\nabla^{\mu}\phi\nabla^{\nu}\phi}{\nabla_{\alpha}\phi\nabla^{\alpha}\phi}\right)\nabla_{\mu}\nabla_{\nu}\phi,
\eeq
to get \eqref{e10} (which is the cuscuton equation of motion). For values of $n\gtrsim \frac{1}{2}$ (slightly greater than $\frac{1}{2}$),
up to  first order in $\epsilon$ \eqref{e8} and \eqref{e9} become
\bea
&\frac{a}{2\sqrt{X}}\left(g^{\mu \nu}-\frac{\nabla^{\mu}\phi\nabla^{\nu}\phi}{2X}\right)\nabla_{\mu}\nabla_{\nu}\phi+V'(\phi)=0,\label{e11}\\
&u_{\mu}v^{\mu}=\epsilon. \label{e12}
\eea
Since equation \eqref{e11} is the same as \eqref{e10},  up to first order in $\epsilon$ the field $\phi$ satisfies the cuscuton equation of motion. Also the propagation cone is determined by equation \eqref{e12}. Because of the key role of the cuscuton field in the discussion, we will explain some of its properties; this will help us to solve \eqref{e11}.
\subsection{Cuscuton Characteristics}
Cuscuton is a scalar field with an infinite speed of sound ($n=\frac{1}{2}$). Its energy-momentum tensor can be expressed in the form of a perfect fluid $T_{\mu \nu}=(\rho +p)u_{\mu \nu}-pg_{\mu \nu}$, where
\bea
&\rho=V(\phi),\label{f1}\\
&p=a\sqrt{X}-V(\phi),\label{f2}\\
&u_{\mu}=\frac{\nabla_{\mu}\phi}{\sqrt{\nabla_{\alpha}\phi\nabla^{\alpha}\phi}}\label{e16},
\eea
and it satisfies the equation of motion \eqref{e10}.

Equation \eqref{e16} shows that constant field surfaces are the same as co-moving surfaces, as the field's velocity vector $u^\mu$  is the normal vector to constant field surfaces. Since the mean curvature $K$ (the trace of the extrinsic curvature) of constant field surfaces is the divergence of the normal vector to the surface, we have
\beq\label{e17}
K= u^{\alpha}_{\phantom{a};\alpha} = -\frac{\sqrt{2}}{a}V'(\phi),
\eeq
where we have used \eqref{e10} in the second equality. Equation \eqref{e17} shows that mean curvature $K$ is constant on a constant field surface. It means that constant cuscuton field surfaces   are CMC (constant mean curvature) surfaces. Consequently in order to find constant cuscuton field surfaces  we only need to find the CMC surfaces of the background spacetime.

\section{CMC Surfaces of Spherically Collapsing Thin Shell of Dust Spacetime}
As we mentioned in the previous section, we only need to find CMC surfaces of the background spacetime to determine the propagation cone. We consider here a  collapsing shell of spherically symmetric dust as the background spacetime and derive its CMC surfaces.

Assuming that the thin shell is located at $r=R(t)$, it divides spacetime into two regions with the following metrics:
\bea
\text{I}&:& ds^2=A^2(t)dt^2-dr^2-r^2d\Omega^2, \text{   $r<R(t)$}\notag\\
\text{II}&:& ds^2=f(r) dt^2-\frac{dr^2}{f(r)}-r^2d\Omega^2, \text{   $r>R(t)$},\notag
\eea
in which $f(r)=1-\frac{2M}{r}$, and we have ignored the gravitational back reaction of the cuscuton field.
The shell radius satisfies the following geodesic equation
\beq\label{e26}
\dot R= -f(R)\sqrt{1-\frac{f(R)}{e^2}},
\eeq
where $e$ is a constant of motion and $\cdot=\frac{d}{dt}$. The function $A(t)$ can be found by matching the line elements at $r=R(t)$
\bea
A=f(R)\sqrt{1+\frac{2M}{e^2R}}.
\eea
In order to find CMC surfaces in this spacetime, we need to find CMC surfaces in each region and match them at $r=R(t)$. If $t_{CMC}=T(r)$ is a CMC surface with constant mean curvature $K$, the normal vector $u_{\mu}$ (in region II) to this surface will be
\beq
u_{\mu}= \frac{1}{N}\nabla_{\mu}(t-T(r))=\frac{1}{N}(1,-T'(r),0,0),
\eeq
where $N$ is the normalization factor 
\beq
N^2=\frac{1}{f(r)}-f(r)T'^2(r)
\eeq
that we choose   to be positive. As a result 
\beq\label{e18}
u^{\mu}=\frac{1}{N}(\frac{1}{f(r)},f(r)T'(r),0,0)
\eeq
and $K=\nabla_{\mu}u^{\mu}$ yields
\beq\label{e19}
u^{r}=\frac{K}{3}r-\frac{B}{r^2},
\eeq
where $B$ is an integration constant. This constant may vary from one CMC surface to another. Comparing \eqref{e18} with \eqref{e19} yields
\beq\label{e20}
T'(r)=\frac{u^r}{f(r)\sqrt{f(r)+\left(u^r\right)^2}}=\frac{\frac{K}{3}r-\frac{B}{r^2}}{f(r)\sqrt{f(r)+\left(\frac{K}{3}r-\frac{B}{r^2}\right)^2}},
\eeq
with the following unit normal  to the CMC surface
\bea
u^r_{II}&=&\frac{K}{3}r-\frac{B}{r^2},\label{e21}\\
u^t_{II}&=\frac{1}{f(r)}&\sqrt{(\frac{K}{3}r-\frac{B}{r^2})^2+f(r)}.\label{e22}
\eea
Similar calculations yield the following unit normal 
\bea
u^r_{I}&=&\frac{K}{3}r,\label{e23}\\
u^t_{I}&=&\frac{\sqrt{(\frac{Kr}{3})^2+1}}{A}. \label{e24}
\eea
to the CMC surface in the first region.

\subsection{Finding $B$}
Our next task is to obtain $B$. Its value can be fixed by matching the two CMC solutions on the shell's surface. We construct a set of  orthonormal basis vectors $\{n,e_i\}$, $i \in \{1,2,3\}$, where the $e_i$'s form a complete orthonormal basis for the shell's surface and $n$ is the unit vector normal to the surface.

The value of $B$ must be chosen such that $K=\nabla_{\mu} u^{\mu}$ remains non-singular on the surface of the shell. The previous derivations for $u$ in region I and II are only valid inside each region and not on the surface. 

\begin{figure}[h]
\begin{center}
\includegraphics[width=0.95\linewidth]{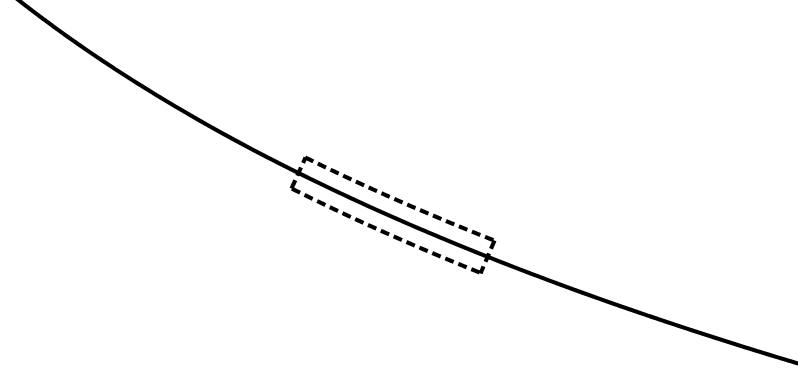}
\end{center}
\caption{Space-time diagram showing $r=R(t)$. The region inside the dashed lines is $V$. Two sides of this region, normal to shell's surface, are much smaller than the other sides, so their contribution to the R.H.S of \eqref{f10} is negligible.}
\label{fig:K}
\end{figure}

However, non-singularity of $K$ can be imposed by using Gauss's law. Consider a small space-time volume $V$ (Fig. \ref{fig:K}); using Gauss's law we find
\beq\label{f10}
\int_V K ~dV = \int_V \nabla_{\mu}u^{\mu}~dV=\int_{\partial V} u_{\mu}n_V^{\mu}~dS,
\eeq 
where $n_V$ is the normal vector to the region $V$.

For sufficiently small $V$  only sides parallel to the shell's surface contribute to the right-hand side of \eqref{f10} and $n_V=\pm n$.  The left-hand side approaches zero as $V\to 0$. Hence $u\cdot n\equiv u_{\mu}n^{\mu}$ must remain continuous across the surface.  Since $K$ is non-singular, we find that $u_I \cdot n=u_{II} \cdot n$, where the equality must be imposed on the shell.


Moreover, in order to have a smooth CMC surface, we demand that the projection of $u$ onto the surface of  the shell   remain continuous, implying  $u_I\cdot e_i=u_{II}\cdot e_i$. Although this smoothness condition results in three equations, two of them are trivial because of spherical symmetry. The non-trivial equation reads
\bea
&g_{tt}u^t_I dt+g_{rr}u^r_IdR=g_{tt}u^t_{II}dt+g_{rr}u^r_{II}dR\label{e25}
\eea
This equation can be solved (analytically) for $B$ in terms of $R$ and $K$, and it has two different solutions. The previous condition (non-singularity) picks one of them. 

It can be easily shown that imposing smoothness condition requires that $u\cdot n$ either remains continuous or flips sign across shell's surface. The correct value of $B$ is the one that does not change the sign of $u \cdot n$.

In the following, we will explicitly derive CMC surfaces in two cases.

\begin{figure}[t]
\includegraphics[width=0.5\textwidth]{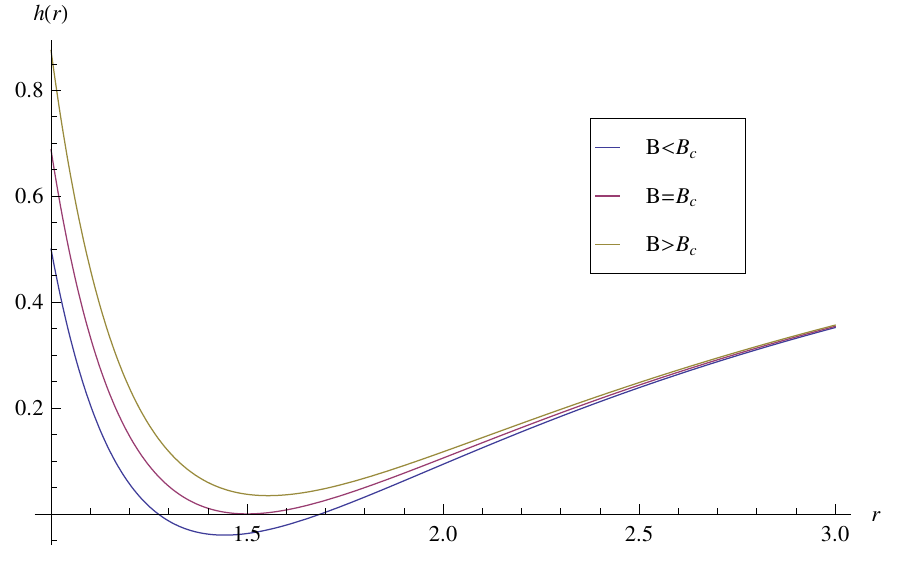}
\caption{The function $h(r,R)$ in terms of $r$ for different values of $B$ (setting $M=1$). $h$ is always positive when $B>B_c$. It has double root at $r=1.5M$ when $B=B_c$ and becomes negative for $B<B_c$.}
\label{fig:h(r)}
\end{figure}

\subsection{$K>0$}
Using \eqref{e26} and \eqref{e20}, we get
\bea
t_{CMC}(r)&=&L(K)\notag\\
&-&\int_r^{K^{-1}} \mathrm{d}x \frac{\frac{K}{3} x -\frac{B}{x^2}}{f(x)\sqrt{f(x)+\left(\frac{K}{3} x -\frac{B}{x^2}\right)^2}},\label{e27}\\
t_{shell}(r)&=&-\int_{r_0}^{r} \mathrm{d} x\frac{1}{f(x)\sqrt{1-\frac{f(x)}{e^2}}}+t_0,\label{e28}
\eea
where $t_0$ and $r_0$ are constants (determining the initial position of shell), $L(K)$ is another integration constant (determining the behavior of CMC surfaces at large radii) and \eqref{e27} is only valid for $r>R$. Note that $t=t_{shell}(r)$ and $r=R(t)$  describe the same surface.

$L(K)$ can be fixed by matching cuscuton solutions to cosmological ones at large distances, remembering that a CMC surface is also a cuscuton constant density surface. Taking the derivative with respect to $K$ in \eqref{e27}, we obtain
\bea
\frac{1}{\dot K} \equiv \left(\frac{\partial t}{\partial K}\right)_{r\sim K^{-1}} &\approx& L'(K)+\frac{1}{K^2}\frac{\frac{1}{3} - K^2 B}{\sqrt{1+(\frac{1}{3} -K^2 B)^2}}\notag\\
&\approx& L'(K)+\frac{1}{K^2}\frac{1}{\sqrt{10}}\label{e29}
\eea
where we have used $MK \ll 1$ (as the Schwarzschild horizon $2M$ is much smaller than the cosmological horizon $K^{-1}$) and $K^2 B \ll 1$ (since we expect to have a homogeneous cuscuton field on cosmological scales \eqref{e19}). Knowing $\dot K$ from cosmology, we can fix $L(K)$.

Once all the constants ($t_0$, $r_0$ and $L(K)$) are fixed, we require $t_{CMC}(R)=t_{shell}(R)$. This equation together with \eqref{e25} for each value of $R$ gives the corresponding value of $K$ and $B$ and completely fixes the CMC surfaces.
Equation
\eqref{e27} can be expressed also in the following form
\bea\label{z1}
t_{CMC}(r)&=&t_{shell}(R)\\
&+&\int_{R}^{r}\mathrm{d} x \frac{\frac{K}{3} x -\frac{B}{x^2}}{f(x)\sqrt{f(x)+\left(\frac{K}{3} x -\frac{B}{x^2}\right)^2}},~r>R\notag
\eea

\begin{figure}[t]
\centering
\includegraphics[width=0.5\textwidth]{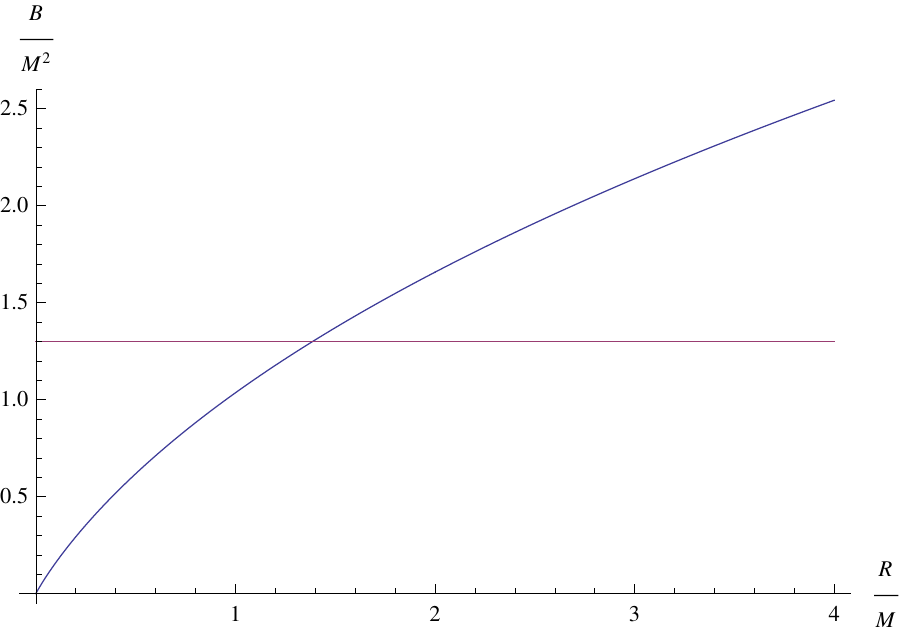}
\caption{$B$ as a function of $R$ for $K=0$ and $e=1$. The horizontal line shows $B=B_c$. The radius at which $B(R)=B_c$ is called $R_c$. Note that $R_c<1.5M$.}
\label{fig:B}
\end{figure}

Notice that \eqref{z1} is a meaningful equation only if for all $r>R$, $h(r,R)\equiv f(r)+\left(\frac{K}{3} r -\frac{B(R)}{r^2}\right)^2$ remains positive ($R$ only fixes the value of $B$). 
Clearly, this condition is satisfied for $r>2M$. For $r<2M$, if we neglect $\frac{K}{3}r$ term (as it is much smaller than $\frac{B}{r^2}$), there is a critical value $B_c=\frac{\sqrt{27}}{4}M^2$ above which $h(r)$ is always positive. For $B=B_c$, function $h(r)$ has a double root at $r=\frac{3}{2}M$ (Fig. \ref{fig:h(r)}). This argument shows that the behavior of CMC surfaces depend heavily on how $B$ changes with $R$ (Fig. \ref{fig:B}).

\subsection{$K=0$}
In this case, we get
\beq\label{e30}
t_{CMC}(r)=-\int_{r'_0}^{r}\mathrm{d} x \frac{\frac{B}{x^2}}{f(x)\sqrt{f(x)+\frac{B^2}{x^4}}}+t'_0 , \mbox{    $r>R$},
\eeq
where $r'_0$ and $t'_0$ are constants. In the case of maximal surfaces ($K=0$), \eqref{e25} gives $B=B(R)$ (for $K=0$), and \eqref{e30} together with $t_{CMC}(R)=t_{shell}(R)$ determine the CMC surfaces for radii larger than $R$. Consequently,
\beq\label{f11}
t_{CMC}(r)=-\int_{R}^{r}\mathrm{d} x \frac{\frac{B(R)}{x^2}}{f(x)\sqrt{f(x)+\frac{B(R)^2}{x^4}}}+t_{shell}(R) , \mbox{    $r>R$},
\eeq

\section{Emergence of the Universal Horizon}\label{emergence}
As we mentioned earlier, an observer inside $r=1.5M$ cannot send any signal outside this radius, after some stage in collapse, even using superluminal $\phi$-signals that propagate almost instantaneously. We shall demonstrate this for two cases.

\subsection{$V(\phi)=0$}
If we set $V(\phi)=0$ in \eqref{e17}, we get
\beq
K= u^{\alpha}_{\phantom{a};\alpha}=0,
\eeq
implying that  constant field surfaces are maximal surfaces. As we showed earlier, in order to determine the signal propagation in this background, we need to find normal vector $u^{\mu}$ to constant field surfaces. Then Equation \eqref{e12} determines the influence cone at any point of spacetime. Using \eqref{e21} and \eqref{e22}, we get
\bea
u^r_{II}&=&-\frac{B}{r^2},\label{e31}\\
u^t_{II}&=\frac{1}{f(r)}&\sqrt{\frac{B^2}{r^4}+f(r)},\label{e32}
\eea
where $B$ is given by \eqref{e25} ($K=0$). As   shown in Appendix \ref{B(R,K)}, there is always a radius $R_c \le1.5M$ for which the corresponding value of $B$ is $B_c$, $B(R_c)=B_c$. It means that when shell's radius approaches $R_c$, the value of $B$   becomes closer to $B_c$, and the t-component of the normal vector ($u^t_{II}$) at $r=1.5M$ approaches zero. Note that $R_c$ must be smaller than $1.5M$; otherwise the $t$-component of the normal vector would be $u^t_I$. 

On the other hand, equation \eqref{e12} at $r=1.5M$ yields
\beq
-\frac{1}{3}u^t_{II}v^t+3u^r_{II}v^r=\epsilon.
\eeq
Consequently, (for a fixed value of $\epsilon$) when $R$ reaches $R_c$ the first term in the above equation becomes negligible. As a result $v^r<0$ (because $B>0$) and the propagation cone becomes tilted toward the center. As a result, no signal can escape $r \leq 1.5M$. The surface $r=1.5M$ is the ``Universal Horizon" as no signal (even with infinite  propagation speed) can escape from within.

\begin{figure}
	\begin{subfigure}[t]{0.5\textwidth}
        		\includegraphics[width=\hsize]{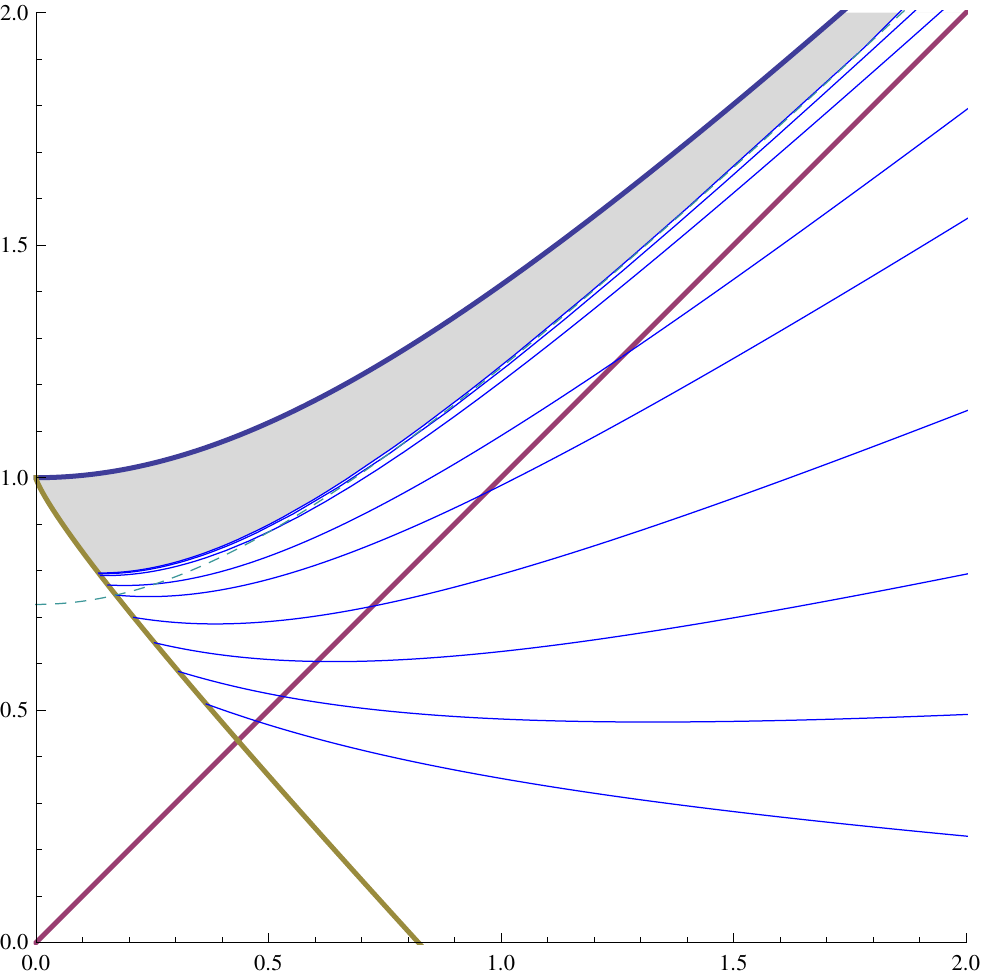}
		\caption{Constant field surfaces in Kruskal coordinates. Thick blue, yellow and brown curves respectively represent $r=0$, the shell's surface and $r=2M$. Blue curves  represent constant field surfaces and the dotted green curve is $r=1.5M$. We see that after some point constant field surfaces tend to stay close to $r=1.5M$.}		
    	\end{subfigure} %
	\hfill
	\begin{subfigure}[t]{0.5\textwidth}
        		\includegraphics[width=\hsize]{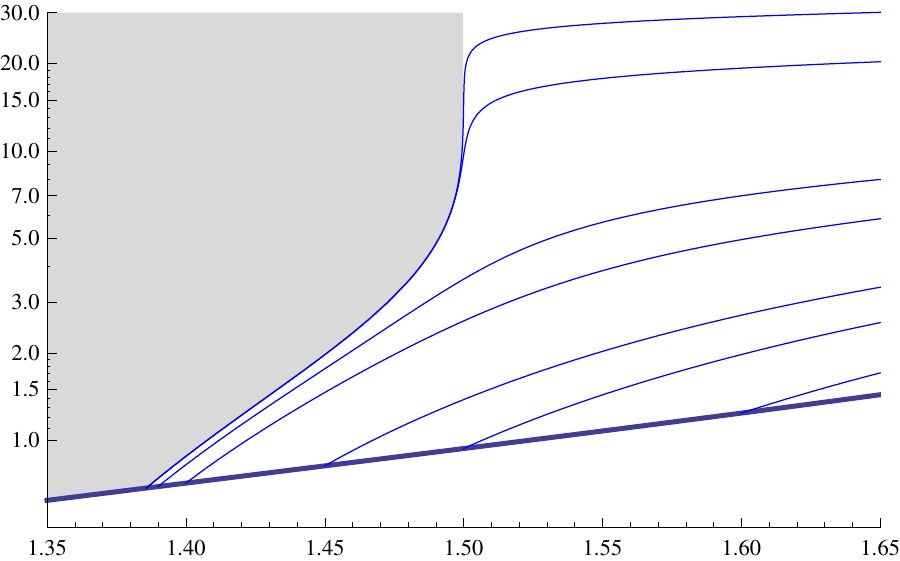}
		\caption{Constant field surfaces in Schwarzschild coordinates. The thick blue curve is the shell's surface, and blue curves are constant field surfaces. We see that after some point they tend to stay close to $r=1.5M$.}
    	\end{subfigure} %
	\caption{Constant field surfaces for $e=1$ and $M=1$ in Kruskal and Schwarzschild coordinates. Grey area shows the region causally disconnected from infinity.}
	\label{fig:B1}
\end{figure}

Maximal surfaces (surfaces of constant field) have been shown in Fig. \ref{fig:B1} in Schwarzschild and Kruskal\footnote{Kruskal coordinates: 
\bea
v&=&\left|\frac{r}{2M}-1\right|^{\frac{1}{2}}e^{\frac{r}{4M}}\left[\sinh (\frac{t}{4M}) \theta(r-2M)+\cosh (\frac{t}{4M}) \theta(2M-r)\right],\notag\\
u&=&\left|\frac{r}{2M}-1\right|^{\frac{1}{2}}e^{\frac{r}{4M}}\left[\cosh (\frac{t}{4M}) \theta(r-2M)+\sinh (\frac{t}{4M}) \theta(2M-r)\right],\notag
\eea } coordinates. Close to $R_c$, maximal surfaces tend to stay very close to $r=1.5M$.




\subsection{$V(\phi)\neq 0$}
The case $V(\phi) \neq 0$   is almost the same as $V(\phi)=0$, as long as $MK \ll 1$.  
This can be argued as follows. We are interested in the regions where the shell radius is of order  $M$. As a result, solutions to \eqref{e25}, as long as $MK \ll 1$, are the same as $K=0$.  Hence a small non-zero value of $K$ will not change the shape of the CMC surfaces at small radii (radii of the order of $M$), and it only affects the shape of CMC surfaces at large radii (cosmological scale). However, our derivation of the universal horizon (in the previous section) only depends on maximal surfaces inside Schwarzschild radius. Consequently, we can apply the same argument to a small non-zero value of $K$.

\subsection{Is the Universal Horizon Singular?}
Until now, we have ignored the effect of the cuscuton on the background geometry. This assumption is valid if the cuscuton's pressure and density remain small. 

Consider a quadratic potential $V(\phi)=\lambda \phi^2$ for the cuscuton. Using the cuscuton's EOM \eqref{e17} we find $K \propto \phi$, implying that the field value is suppressed by the Hubble parameter ($K=3H$ where $H$ is Hubble parameter) . As a result the cuscuton's energy density $\rho=V(\phi)$ is also Hubble-suppressed. Since the cuscuton pressure is given by the time derivative of $\phi$ \eqref{f2}, it is suppressed by $\dot H$. This argument shows that the cuscuton energy density and pressure are small. However, since constant field surfaces tend to stack around $r=1.5M$ upon formation of  the universal horizon, the above argument may not be applicable  in this limit. We show here that upon forming a universal horizon, this surface is non-singular. In fact the cuscuton's pressure and density remain small when shell radius approaches $R_c$.

According to \eqref{f1} the density always remains finite for a well-behaved potential and boundary condition. As an example, for $V(\phi)=\lambda \phi^2$ and matter dominated cosmology, $\rho$ remains small (in fact, it decreases with time and approaches zero).
Since the pressure is given by \eqref{f2}
\beq
p=a\sqrt{X}-V(\phi),
\eeq
we need to show that $2X=g^{\mu \nu}\partial_{\mu}\phi \partial_{\nu}\phi$ remains finite in the limit $R \rightarrow R_c$.

In the following we  assume that $V(\phi)$ is at least quadratic in $\phi$.
Differentiating \eqref{e17}, we get
\beq
\partial_{\mu}K=-\frac{\sqrt{2}}{a}V''(\phi)\partial_{\mu}\phi.
\eeq
As a result, 
\bea
2X&=&\frac{a^2}{2\left(V''(\phi)\right)^2}g^{\mu \nu}\partial_{\mu}K\partial_{\nu}K\notag\\
&=&\frac{a^2}{2\left(V''(\phi)\right)^2}\left[\frac{1}{f(r)}\left(\partial_tK\right)^2-f(r)\left(\partial_rK\right)^2 \right].\label{f3}
\eea
Using the identity
\beq
\left(\frac{\partial K}{\partial r}\right)_t=-\left(\frac{\partial t}{\partial r}\right)_K\left(\frac{\partial K}{\partial t}\right)_r,
\eeq
 together with \eqref{e20}, equation \eqref{f3} yields
\beq
2X=\frac{a^2}{2\left(V''(\phi)\right)^2}\frac{\left(\frac{\partial K}{\partial t}\right)_r^2}{f(r)+\left(\frac{K}{3}r-\frac{B}{r^2}\right)^2}.
\eeq
In order to have a singularity the following term 
\beq\label{f4}
\left[f(r)+\left(\frac{K}{3}r-\frac{B}{r^2}\right)^2\right]\left(\frac{\partial t}{\partial K}\right)_r^2
\eeq
must approach zero. Let us provisionally assume that
\beq\label{f5}
\left(\frac{\partial t}{\partial K}\right)_r =\frac{1}{\dot K}-\int_r^{K^{-1}}\mathrm{d} x \frac{\frac{x}{3}-\frac{dB/dK}{x^2}}{\left[f(x)+\left(\frac{K}{3}x-\frac{B}{x^2}\right)^2\right]^{\frac{3}{2}}}
\eeq
does not approach zero. The other term $f(r)+\left(\frac{K}{3}r-\frac{B}{r^2}\right)^2$ can reach zero upon formation of universal horizon ($R\rightarrow R_c$, $B\rightarrow B_c$) at $r=1.5M$. However, $\left(\frac{\partial t}{\partial K}\right)_{r=1.5M}$ is also diverging at the same limit. Considering \eqref{f4} in the limit $R\rightarrow R_c$ and $r\rightarrow 1.5M$ we find
\bea
&~&\left[f(r)+\left(\frac{K}{3}r-\frac{B}{r^2}\right)^2\right]\left(\frac{\partial t}{\partial K}\right)_r^2\notag\\
&~&\sim(r-1.5M)^2\times \frac{1}{(r-1.5M)^4} \sim \frac{1}{(r-1.5M)^2}\label{f6}
\eea
showing that not only \eqref{f4} does not reach zero,  it rather diverges as $r\rightarrow 1.5M$. Hence $\rho=-p$ at $r=1.5M$ in the limit of the formation of the universal horizon.

The behaviour of $\left(\frac{\partial t}{\partial K}\right)_r$ heavily depends on the value of $dB/dK$. Note that if $\left(\frac{\partial t}{\partial K}\right)_r$ becomes zero at some point, it means that two CMC surfaces crossed each other. As a result, it seems that as long as we are able to find solutions for the cuscuton, $\left(\frac{\partial t}{\partial K}\right)_r$ never vanishes. In the following, we will prove (by contradiction) that for $\dot K<0$, this term does not vanish.

If $\left(\frac{\partial t}{\partial K}\right)_r=0$ at some point, $dB/dK$ must be positive; otherwise the integrand in \eqref{f5} will be always positive and $\left(\frac{\partial t}{\partial K}\right)_r<0$.

\begin{figure}[h]
\begin{center}
\includegraphics[width=0.95\linewidth]{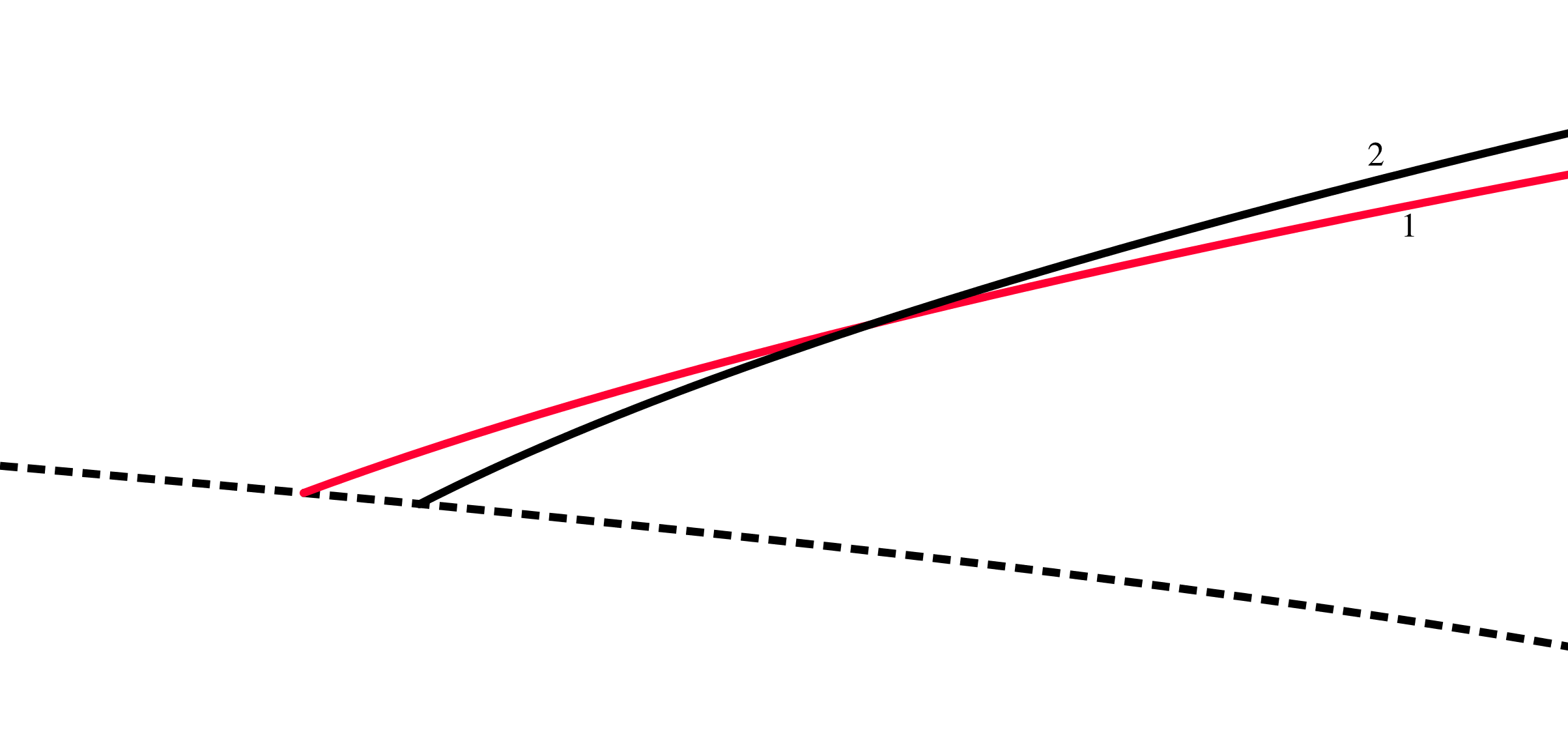}
\end{center}
\caption{Space-time diagram showing two crossing CMC surfaces and the shell's surface (dashed line).}
\label{fig:CrossingCMC2}
\end{figure} 

On the other hand, $\left(\frac{\partial t}{\partial K}\right)_r=0$ means that two CMC surface have crossed each other (as in Fig. \ref{fig:CrossingCMC2}). 
Note that these surfaces must cross and cannot be tangent to each other because $\left(\frac{\partial t}{\partial K}\right)_r$ changes sign near its zero. It means that $R_1\le R_2$ (where $R_i$ is the shell radius corresponding to  surface $i$), and consequently $B(R_1) \le B(R_2)$, valid at least when the shell's radius is close to $R_c$.

If we assume that these surfaces are infinitesimally close to each other, then
\beq
dB/dK=\frac{B(R_1)-B(R_2)}{K_1-K_2}\le 0.
\eeq
Note that the crossing point can be on the shell, which in this case $dB/dK=0$. This is in contradiction with the fact that $dB/dK$ must be positive. As a result, we have shown that for negative values of $\dot K$, $\left(\frac{\partial t}{\partial K}\right)$ never becomes zero.

In Appendix \ref{limit}, we have  derived an approximate formula for $dB/dK$ which further shows that $\left(\frac{\partial t}{\partial K}\right)$ does not become zero.

\section{Summary and Conclusions} \label{conclude}
Constant field hypersurfaces of a cuscuton field represent the preferred time slicing in the low energy limit of Ho\v{r}ava-Lifshitz gravity. In this paper, we have demonstrated that, although the space-time geometry of a large black hole in Ho\v{r}ava-Lifshitz gravity is very similar to Schwarzschild geometry, the causal structure is completely different.

Nevertheless, we showed that as a black hole forms, there still exists an event horizon for signals with arbitrarily large speed. No matter how fast a signal propagates, it  cannot escape from inside this {\it universal horizon} (i.e. grey area in Fig. \ref{fig:B1}), as seen in static solutions previously \cite{Blas:2011ni, Barausse:2011pu, Barausse:2012qh}. If this was not the case, one could have imagined that signals originating from regions close to the singularity could in principle propagate outside, leading to a {\it naked singularity}, and rendering the classical theory unpredictive. Instead, the emergence of a universal horizon during the formation of the black hole implies that a version of {\it Cosmic Censorship} might still hold here. 
These results are also consistent with earlier studies of the gravitational collapse of K-essence matter \cite{Akhoury:2011hr,Leonard:2011ce}, in which sonic horizons could form inside luminal horizons, and  gravitational collapse in the context of Einstein-aether theories \cite{Garfinkle:2007bk}.

 This causal structure has an additional interesting property. The universal horizon relates the finite time coordinate inside black hole to  future infinity (outside). As a result, any observer falling into black hole will hit the universal horizon at a finite proper time, prior to which she can, in principle, see events that happen outside black hole at arbitrarily late times (if she can see arbitrarily superluminal signals in the preferred frame). 
For example, the black hole itself radiates its mass through Hawking evaporation, with quantum mechanical effects becoming important when the black hole radiates a substantive portion of its mass.   While, for an observer outside a  massive   black hole, this takes a long time ($\sim M^3$), an observer falling into black hole can detect these quantum mechanical effects (via superluminal contact with outside) within a much shorter time ($\sim M$), just before hitting the universal horizon. In this sense, the universal horizon can be considered the causal future boundary of the classical space-time. 

This realization could also be intimately related to the claim that, while the universal horizon in our spherically symmetric system is regular, it is unstable to aspherical perturbations   that  might change it to a singular surface \cite{Blas:2011ni}.  
On the other hand,  \cite{Berglund:2012bu, Berglund:2012fk} argue that, similar to the ordinary null horizons, universal horizons may radiate particles, and  a fixed temperature and entropy can be assigned to them. However, it seems that the instability of universal horizon by aspherical perturbations is incompatible with the derivation of horizon temperature. 

Another important issue regarding Lorentz violating theories is their apparent tension with generalized 2nd law. It has been argued that in a theory with two different fields $A$ and $B$ with different speeds $c_A$ and $c_B$ where each has its own horizon around black hole, one can violate generalized 2nd law with building a perpetual motion machine (see for example \cite{Dubovsky:2006vk,Eling:2007qd}).
Clearly, there remains a lot to be understood about the nature of black holes in Lorentz violating theories, and their {\it universal horizons}.

\vspace*{10mm}

{\it Acknowledgement:} The authors would like to thank Siavash Aslanbeigi, Rafael Sorkin, Tim Koslowski, David Mattingly, Diego Blas and Enrico Barausse for invaluable discussions. This work was supported by the Natural Science and Engineering Research Council of Canada, the University of Waterloo and the Perimeter Institute for Theoretical Physics. Research at Perimeter Institute is supported by the Government of Canada through Industry Canada and by the Province of Ontario through the Ministry of Research \& Innovation. 

\bibliographystyle{ieeetr}
\bibliography{Universal-Horizon_v8}

\appendix
\pagebreak
\section{Proof of the Existence of Universal Horizon}\label{B(R,K)}

Here, we investigate the conditions for the existence of a universal horizon for $K=0$ but arbitrary value of $e$.
As we showed in section \ref{emergence}, a universal horizon will appear if there is a radius $R_c$ smaller than $1.5M$ such that the corresponding value of $B$ is $B_c$. In fact, this is one of the conditions for the existence of universal horizon.
\\

{\it Condition} 1:\\
There exists $ R_c \le 1.5M$ such that $B(R_c)=B_c$.\\

We prove that this condition is always satisfied. Solving \eqref{e25} for small values of $R$, we find  $B(R=0)=0$. It can be checked easily that the minimum value of $B$ at $R=1.5M$ is always greater than or equal to $B_c$. Hence there must be $R_c \le 1.5M$ such that $B(R_c)=B_c$. If there is more than one solution to $B(R)=B_c$ for $R \le 1.5M$, we call the biggest one $R_c$.

One more condition also needs to be satisfied: solutions of CMC surfaces must be well defined before the shell radius reaches $R_c$. In other words, solutions in region II must be well defined all the way to $R_c$.
\\

{\it Condition} 2:\\
There does not exist $R>R_c$ such that for some value $r>R$, $h(r,R)<0$.\\

In the proof of condition 1, we showed that for $ R_c \le R \le 1.5M$, $B(R)\ge B_c$ (otherwise there is a violation of the condition that $R_c$ is the biggest root of $B(R)=B_c$, which is smaller than $1.5M$). This means that $h(r,R)$ is always positive for $R_c \le R \le 1.5 M$ (Figure \ref{fig:h(r)}). As a result, condition 2 is satisfied for this range of $R$. Furthermore it is obvious that condition 2 is satisfied for $R>2M$. 

Finally, we prove that the condition 2 is satisfied for region $1.5M< R<2M$ by contradiction.
Assume that a radius  $R_0>1.5M$ exists such that for some value $r_0>R_0$, $h(r_0,R_0)<0$. By the properties of the function $h$ it is clear that $B(R_0)<B_c$ (otherwise $h$ is always positive).
Also, $h(r,R_0)$ has only one {\it minimum} at $r_{min}=(\frac{2B^2(R_0)}{m})^{1/3}$. Using $B(R_0)<B_c$, it is clear that $r_{min}<1.5M$. Since the function $h$ has only one minimum, we conclude 
\beq
r_{min}<1.5M<R_0<r_0 \rightarrow h(R_0,R_0)<h(r_0,R_0)<0.
\eeq
However, it can be checked directly that $h(R,R)=1-\frac{2m}{R}+\frac{B^2(R)}{R^4} \ge 0$.

Consequently, the appearance of a universal horizon has been shown for arbitrary value of $e$.
\newpage
\section{Dependence of $B$ on $K$}\label{limit}
Here, we want to derive an analytic expression relating $B$ to $K$ in the limit of universal horizon formation. We will use the following approximation for this derivation.

First, we match CMC surfaces to cosmological ones deep inside the Hubble radius. This approximation let us  use \eqref{f11} to find a time coordinate for each CMC surface at large radii (large compared to the Schwarzschid radius and small compared to the cosmological horizon). Then, knowing $K=K(t)$ for a specific cosmology, we can assign a value of $K$ to each surface.
Second, we perform all calculations in the limit $R\rightarrow R_c$. In this limit the time coordinate of each CMC surfaces goes to infinity. We are interested in the leading order divergent term. In the following, when two equations are related through $\sim$, it means that they are equivalent up to their leading order.

We begin by finding $t_{CMC}$. According to \eqref{f11}, we have
\bea
&~&t_{CMC}(r)=t_{shell}(R)-\int_{R}^{r}\mathrm{d} x \frac{\frac{B}{x^2}}{f(x)\sqrt{f(x)+\frac{B^2}{x^4}}}\notag\\
&=&t_{shell}(R)-M\int_{R/M}^{r/M}\mathrm{d} x \frac{b x}{\left(x-2\right)\sqrt{x^4-2x^3+b^2}}\notag\\
&=&t_{shell}(R)-\notag\\
&~&M\int_{R/M}^{r/M}\mathrm{d}x \frac{b x}{\left(x-2\right)(\left(x-1.5\right)^2\left(x^2+x+\frac{3}{4}\right)+b^2-b_c^2)^{1/2}}\notag
\eea
where $b \equiv \frac{B}{M^2}$ and $b_c\equiv \frac{B_c}{M^2}=\frac{\sqrt{27}}{4}$.

In the limit $R\rightarrow R_c$ ($b\rightarrow b_c$), the divergent term in the last equation comes from the integral around $x=1.5$. Considering that $t_{shell}(R)$ limits to a constant value, we find
\bea
&~&t_{CMC}(r) \sim \notag\\
&-&M\int_{1.5-\epsilon}^{1.5+\epsilon}\mathrm{d} x \frac{b x}{\left(x-2\right)\sqrt{\left(x-1.5\right)^2\left(x^2+x+\frac{3}{4}\right)+b^2-b_c^2}}\notag\\
&\sim&3Mb_c\int_{1.5-\epsilon}^{1.5+\epsilon}\mathrm{d} x \frac{1}{\sqrt{\frac{9}{2}\left(x-1.5\right)^2+b^2-b_c^2}}.\label{appb2}
\eea
Using the following identity for small $y$
\beq
\int_{-\epsilon}^{+\epsilon}\frac{\mathrm{d}x}{\sqrt{z^2x^2+y^2}}\sim -\frac{\ln(y^2)}{|z|},
\eeq
\eqref{appb2} yields
\beq
t_{CMC}\sim -\sqrt{2}Mb_c\ln\left(b^2-b_c^2\right).
\eeq

For a specific cosmological scenario, we can relate $K$ to $B$ through $K=K\left(t_{CMC}\right)$. 
As an example, for a matter dominated cosmology $K=\frac{2}{t}$, and so
\beq
K=\frac{2}{t_{CMC}}\sim -\frac{\sqrt{2}}{Mb_c\ln(b^2-b_c^2)},
\eeq
which results in 
\beq
\frac{dK}{dB}=\frac{1}{M^2}\frac{dK}{db}=\frac{2\sqrt{2}}{b_cM^3}\frac{b}{\left(b^2-b_c^2\right)\left[\ln\left(b^2-b_c^2\right)\right]^2}.
\eeq
As a result, $dK/dB\rightarrow \infty$ as $b\rightarrow b_c$. Consequently  $\left(\frac{\partial t}{\partial K}\right)_r$ does not approach zero in this limit.
As another example,   consider $\Lambda$CDM cosmology. At late times, we have 
\beq
K^2=K_{\Lambda}^2+K_0^2e^{-K_{\Lambda}t_{CMC}}\sim K_{\Lambda}^2+K_0^2\left(b^2-b_c^2\right)^{\sqrt{2}b_cMK_{\Lambda}},
\eeq
which again gives $dK/dB \rightarrow \infty$ ($MK_{\Lambda} \ll 1$).

\end{document}